# Public Health, Technology, and Human Rights: Lessons Learned from Digital Contact Tracing

Maria Carnovale, Khahlil Louisy

## Abstract

To mitigate inefficiencies in manual contact tracing processes, Digital Contact Tracing and Exposure Notifications Systems were developed for use as public-interest technologies during the SARS-CoV-2 (COVID-19) global pandemic. Effective implementation of these tools requires alignment across several factors, including local regulations and policies and trust in government and public health officials. Careful consideration should also be made to minimize any potential conflicts with existing processes in public health which has demonstrated effectiveness. Four unique cases—of Ireland, Guayaquil (Ecuador), Haiti, and the Philippines—detailed in this paper will highlight the importance of upholding the principles of Scientific Validity, Necessity, Time Boundedness, and Proportionality.

**Table of Contents**

# Executive Summary

The devastating toll on human lives consequent to the SARS-CoV-2 global pandemic underscored a pressing need for innovations in public health practice to respond to modern-day crises at the population level. The exponential rise in incidence and prevalence rates of the disease between March 2020 when it was declared a pandemic by the World Health Organization and the proceeding months, pressured governments and health practitioners to seek alternative solutions to existing processes or to augment their practices. Several technology companies and tech-driven non-profit organizations developed digital tools to assist contact tracers with their work, each claiming that their solutions were privacy-preserving, would improve efficiency in the contact tracing process, and benefit marginalized and vulnerable communities.

Digital Contact Tracing and Exposure Notification Systems were among the most prominent of those tools and were largely driven by Apple and Google. Rather than relying on memory, individuals would download a mobile app and give permission to exchange Bluetooth keys. Should an individual test positive to COVID-19, they would simply upload a verified code given to them by healthcare providers, which would trigger a notification to other devices in the population if they were in close enough proximity to the infected individual for their phones to exchange keys. Other solutions used GPS technology which required constant tracking and logging of the users' location. For those using GPS-reliant apps, they would transfer a file containing a log of their location–latitudinal and longitudinal points over a period of time to the contact tracer, who then sent out the exposure notification.

The extent to which these novel technologies were being implemented globally, triggered a team of researchers at Oxford University to develop a set of guiding principles for ethically implementing digital contact tracing and exposure notification apps. These principles were derived from the European Convention on Human Rights, the International Covenant on Civil and Political Rights (ICCPR), and the United Nations Siracusa Principles. Under these principles, the requirements for implementation of the app were met if there were (i) Scientific Validity—sufficient scientific evidence that health outcomes could be improved through the use of the app; (ii) Necessity—no available alternatives would be less harmful; (iii) Time-Boundedness—provisions for sunsetting the collected data in a realistic time frame; and (iv) Proportionality—the gravity of the situation justified its implementation, despite any social and individual costs.

In this paper, each of the four principles is detailed in a specific case, informed by high-level engagements between various governments, public health departments, or multilateral organizations and the PathCheck Foundation, one of the largest providers of exposure notification systems internationally. The Irish use of NearForm and the Ireland Health Service Executive's COVID Tracker app forms the basis of the Scientific Validity case. A citizen's response in Guayaquil, Ecuador is summarized under Necessity. Transferring individual data from the app to a different setting from its intended use is detailed under Time-Boundedness, and the Philippines' attempt to develop a citizen surveillance tool is reflected upon under Proportionality.

Implementation of digital technologies is non-trivial, even when they have been developed for use in the public's interest. This is at a time when trust in both governments and big tech companies is low. To ease this process, governments and policymakers need to consider the range of possible externalities that could result from these implementations and find appropriate solutions as responses, in a transparent way. Transparency is a necessary ingredient in the adoption process – it builds trust between citizens and the implementors and encourages participation in public discourse. Protections against unintended use of the tools must also be instituted, thereby limiting any potential for violations against human rights.

The effectiveness of any piece of public-interest technology is dependent on the complementarity of the local policies in the context within which it is being deployed. If these digital tools are to be substitutes for public service, their operational processes should not run afoul of existing policies or regulations, nor should these policies hinder the processes which make them effective. To that end, successful implementations involve the two groups—policymakers and technologists—working together to understand the constraints or limitations within which the tools must operate, and be designed to be compatible.

## Introduction

Social distancing, lockdowns, limited travel, and contact tracing, both digital and manual, have been the main ingredients to most countries' public health responses to the COVID-19 pandemic since its onset in late 2019. Even those countries that contained the spread of the virus suffered enormous economic costs. The International Monetary Fund estimated that the pandemic might cost 9 trillion dollars only in output lost across 2020 and 2021 (Gopinath). Countries have been integrating technology in the design of effective public health responses to limit their economic burden and harm to individual rights and liberties. Many of those implementations have been controversial and subject to debates over privacy and violation of other human rights.

This paper outlines a set of guidelines for countries to align their use of public interest technologies with the protection of the human rights of their residents, based on observation of best practices and failed attempts in the use of digital contact tracing and exposure notification apps during the COVID-19 pandemic. The experiences of Ireland, Guayaquil (Ecuador), Haiti, and the Philippines described in this paper will highlight the importance of upholding the principles of scientific validity, necessity, time boundedness, and proportionality. Those cases will outline how violating these standards is socially harmful and might also hamper the effectiveness of the policy intervention by misaligning the incentives to the final users with the objectives of the policymaker.

There were (and still are) many open questions about the most effective policy responses to the COVID-19 pandemic. These questions are partly driven by uncertainties on its underlying virus—SARS-CoV-2. For example, the quick spread of the infection led researchers to posit that, rather than simply transmitting through respiratory droplets (fluids expelled when coughing, sneezing, talking, etc.) as it was initially believed the virus might be airborne—infective via aerosols suspended in air over long distances and time (World Health Organization, *Transmission of SARS-CoV-2: Implications for Infection Prevention Precautions*). In most countries, policy responses were adjusted accordingly. Estimates of fatality rates —the number of COVID-related deaths over the number of positive cases—were severely undermined by the lack of large-scale testing. More recent scientific studies are converging around 0.5%- 1%. For reference, the mortality rate of the common flu is 0.1% (Mallapaty).

SARS-CoV-2 is a Coronavirus, the same virus family of SARS and MERS (National Institutes of Health). It can cause a wide range of symptoms from COVID-19—a debilitating and potentially deadly inflammation of the upper respiratory tract—to mild cough and other flu-like symptoms (Center for Disease Control). Its host is infective for up to 14 days before the onset of symptoms and most people carry the virus asymptomatically (Housen et al.). Therefore, its control requires strict preventive measures.

China, where the virus first emerged, contained it by imposing a severe lockdown and social distancing requirement in affected areas (Illmer et al.; Mei; Ji et al.). New Zealand quickly closed its borders to limit the number of cases in the country (James et al.; Reuters). Singapore deployed an encompassing manual contact tracing program to track infected individuals and quarantine them (Sagar). Most other countries followed their example with a combination of similar public health measures.

In all countries, contact tracing has been a key element of this mix. It was a cornerstone in the response to other virus outbreaks, such as Ebola in 2014 and SARS in 2003 (De La Garza). It is also routinely used to monitor and contain sexually transmitted infections, such as HIV/AIDS (Kass and Gielen) and chlamydia (England et al.), and other communicable diseases like tuberculosis (World Health Organization, "Tuberculosis"; Loredo et al.) and measles (World Health Organization, *Measles*).

Manual contact tracing consists of identifying people who have been in contact with an infected individual and are thus at higher risk of having contracted the virus. Contact tracers—trained individuals working under the direction of public health authorities—ask those who have tested positive to list who they have been in close contact with before the onset of symptoms. Then, they reach out to all those contacts, assess their risk of having contracted the disease, and suggest precautionary measures like monitoring the development of symptoms, getting tested, or quarantine, and ask to disclose their contacts to pursue potential downstream cases (Landman).

Contact tracing might slow the pace of transmission to levels that allow countries to safely lift some of the restrictions on individual movements and gatherings (Hellewell et al.; Keeling et al.). However, at peaks in infections, it requires a massive logistical infrastructure that includes hiring and training of a large number of contact tracers (Peak et al.).

To increase the cost-effectiveness of contact tracing programs and protect individual privacy, some jurisdictions have adopted digital contact tracing and exposure notification technology—smartphone applications that detect proximity to other smartphones. If a user identifies as COVID-positive, the app notifies any user that has been in proximity and suggests containment measures based on the probability of infection as calculated by an algorithm. Some digital contact tracing applications also give the user the option of sharing individual location data with public health authorities.

It is still unclear whether digital contact tracing and exposure notification apps effectively reduce the infection rate. Most countries were struggling to integrate digital solutions with other public health measures. Digital contact tracing and exposure notification have been a contentious technology worldwide for concerns over individual privacy. Short of a few successful implementations like those in Germany, Switzerland, and Ireland, most countries are hovering around 5% adoption rates (MIT Technology Review COVID Tracing Tracker).

In the following sections, we are going to provide a short overview of the technologies that have powered most national digital contact tracing and exposure notification apps followed by an outline of the major benefits and concerns. Then, we will explain the four guiding principles of scientific validity, necessity, time-boundedness, and proportionality that will base our subsequent analysis of the case of Ireland, Guayaquil (Ecuador), Haiti, and the Philippines. Finally, we will conclude with a set of recommendations for governments.

## Digital Contact Tracing vs. Exposure Notifications

Manual contact tracing is a comprehensive, systematic approach used by public health experts to discover the path of an infectious disease from the individual who has been identified as a positive case, all the way back to the source of infection. Once a positive case has been identified, a public health expert, typically referred to as a contact tracer, interviews the patient to recall all interactions they had, as well as public venues they spent significant time in. The threshold for the time duration of these interactions and proximity to others differs by the type of infection. For SARS-CoV-2, the U.S. Center for Disease Control (CDC) defines close contact as "as any individual who was within six feet of an infected person for at least 15 minutes starting from 2 days before illness onset (or, for asymptomatic patients, 2 days before positive specimen collection) until the time the patient is isolated" (Center for Disease Control and Prevention).

Once all contacts have been identified, the contact tracer alerts each of those individuals of the possible exposure as quickly and sensitively as possible and evaluates and monitors them for symptoms of the infection. Contact tracing is often accompanied by case management.

Digital Contact Tracing and Exposure Notification are often conflated terms. They describe a novel approach in public health crises that incorporates the use of technology aimed at increasing efficiency and accuracy in the contact tracing process. Individuals in a community are asked to download a mobile application, which, using either GPS or Bluetooth technology, maintains a log of the locations that the user spent time in and/or estimates proximity to others. Should someone in that community become infected, rather than relying on memory, the mobile application provides the contact tracer with precise locations and times that the infected user was in a particular location if using GPS technology. Both technologies send out notifications to anyone who was within a set of parameters that were programmed into the application being used by that individual.

The two technologies also differ in the type of information that is available to contact tracers. GPS technology provides contact tracers with a mapping of the locations that the positive case spent time in, as well as other variables such as duration of time spent in each location, date of visit, and in some cases, the speed at which the individual traveled between points, giving the contact tracer information about mode of travel. The Bluetooth solutions built on Apple and Google's Exposure Notifications API (Apple) are more privacy-preserving. Those applications simply send out alerts or "notifications" to users, warning of possible exposure. Contact tracers do not have access to the level of information derived from the GPS solutions, though individual users may provide such data, in some applications.

Digital exposure notification was developed to augment manual contact tracing, not replace it. In theory, the use of technologies that keep track of precise locations and proximity to others could be more reliable than individual accounts where memory fails. Of course, what follows receipt of these notifications by users is critical to the containment efforts of health professionals. This is where case management plays a vital role and Public Health professionals must think of how it fits into their overall strategies for containment of the disease.

## The Benefits of Mobile Applications in Contact Tracing

### Tracing transmission via anonymous contact

Manual contact tracing has been pivotal in limiting the transmission of sexually transmittable diseases, such as HIV/AIDS (Saxe), but might not be equally powerful in combatting airborne diseases, such as measles. Viruses that are transmitted through brief and indirect contact, such as by touching a surface previously touched by an infected individual or by sharing the same physical space, multiply the channels of transmission and create the potential for anonymous contacts. Manual contact tracers are ill-equipped against those as an individual that tested positive usually might be only capable of remembering contacts of known individuals or locations (De La Garza).

Since its onset, SARS-CoV-2 gave rise to multiple identified super-spreader events—settings where a COVID-positive individual infects many people, usually by casual contact in public spaces or gatherings (Wong and Collins; Leclerc et al.). Super-spreader events greatly increase the rate of spread of the virus and have been the main motivators for lockdown orders in the first months of the pandemic. Yet, even if manual contact tracers identify those that initiated the spread of the virus, it is unlikely they will be able to recall all those they all have been in touch with and often are only able to share their location. A digital contact tracing or exposure notification system, instead, can record unknown contacts.

### Long incubation periods and the limits of human memory

Studies suggest that 14 days is the incubation period of SARS-CoV-2—the time between infection and the onset of symptoms. For most of those two weeks, the SARS-CoV-2 carriers can transmit the virus to others (Housen et al.). The length of the incubation period also limits the efficacy of manual contact tracers since an individual is more likely to provide an accurate account of the previous four days (the length of period before the onset of visible symptoms when measles is infective (World Health Organization, *Measles*) than two weeks. A digital contact tracing or exposure notification system can support individual memory by automatically recording close contacts.

## Privacy issues

To overcome these drawbacks, South Korea had empowered its manual contact tracers to use tools other than mere interviews to investigate the chains of infection. For instance, to respond to its 2015 Middle East Respiratory Syndrome (MERS), South Korea was using patient interviews together with GPS data, credit card transaction records, and surveillance camera footage, giving rise to privacy issues (De La Garza). Singapore's COVID-19 response included similar measures (Sagar; Lai et al.). The use of these data draws attention to the risk that the pandemic will lead to widespread surveillance and limitations of individual agency.

Even if the interviewed individuals report their whereabouts and interactions voluntarily, they also disclose the identity and location of third parties without their consent (Kass). The privacy of some subgroups, like individuals attending Alcoholics Anonymous meetings, or those attending LGBTQ meetings in conservative communities, might be more sensitive than others and its loss might result in discrimination or social stigma. The possibility to share contacts' data anonymously might improve the ability of public health officials to track the dispersion of a virus. While unfeasible in manual contact tracing, some digital contact tracing and exposure notification apps offer that opportunity.

## Cost of scaling up

There are various estimates of SARS-CoV-2's transmission rate, R (the average number of people an infected individual transmits the virus to). Most place R between 2.5 and 6.6: higher than the seasonal flu's 1.3, but lower than measle's 12 (Beech). At this high rate of contagion, manual contact tracing can only be effective when it can be proportionally scaled up. Hiring and training the many contact tracers necessary to effectively keep track of the spread of the virus is costly, especially if the expectation is to lift constraints on people's movements and gatherings.

Early estimates by the Center for Health Security at Johns Hopkins University indicated that health departments in the US would need around $3.6 billion in emergency funds to deploy effective programs, including an expansion of the public health workforce to assist the large-scale monitoring and reporting of cases and to provide access to testing for those with a high probability of infection (Johns Hopkins Center for Health Security). Digital contact tracing and exposure notification apps, on the other hand, have limited scale-up costs and can seamlessly adjust to peaks and valleys in infections.

## Concerns of Digital Contact Tracing and Exposure Notification Apps.

**Unknown effectiveness**

During the last six months of 2020, in an effort to support manual contact tracers, many jurisdictions have launched digital contact tracing or exposure notification apps even though their effectiveness is still unknown. Researchers at the University of Oxford have estimated that 6.1% of individuals notified by the National Health Services (NHS) exposure notification app subsequently test positive, a number comparable to the 7.3% of manual contact tracing (Abueg et al.). Yet, these results are not final: the study is based on observational data rather than systematic experiments. Therefore, while it provides suggestive evidence of the effect of the NHS app, it cannot rule out the possibility that other factors played into it.

The effectiveness of these digital solutions relies on capacious public health infrastructure. While there is still a lot of uncertainty, around 40% of all cases might be asymptomatic (Oran and Topol; Goodman) and asymptomatic cases might account for half of the virus transmission (Johansson et al.). Without random testing widely available, most of those cases would be missed by both manual contact tracers as well as digital systems, together with the chain of infections they spur.

Both systems are only as effective as the individual response that they trigger. Without proper support and resources for those that are unable to self-isolate; without employment protections; and without widespread healthcare coverage, a simple notification might not lead individuals to contact health authorities. Digital contact tracing systems and exposure notification would not provide public health officials with the information to initiate a contact.

High levels of user download and participation are also necessary for digital solutions to produce the desired effects. Most national contact tracing mobile applications lacked widespread adoption (MIT Technology Review COVID Tracing Tracker).

**Exacerbate existing inequalities**

Adoption rates are lowered by trust and resources. Because of systemic disparities in employment, housing, and healthcare access, in the US racial minorities have been most vulnerable to SARS-CoV-2. Overreliance on digital contact tracing and exposure notification apps is likely to increase that racial gap.

Fear that Bluetooth or location data will be used for purposes other than COVID-19 response might deter overpoliced communities from downloading digital contact tracing apps. Some of these digital solutions, like GAEN (Google and Apple Exposure Notification), used by many jurisdictions as the building block of their government-backed exposure notification app, only works with recent operating systems (iOS 13.5 and Android version 6 or more recent), often not compatible with phones older than 5 years (Reader). For example, Australia had no plan on making the app work with operating systems older than iOS 10 or Android 6 (Taylor).

Low-income individuals, often with older phones, might lack access altogether to digital contact tracing and exposure notification apps. Low-income countries might not be effectively served by this digital innovation because they often have lower levels of smartphone penetration and an overall older smartphone fleet.

**Data privacy**

Most digital contact tracing and exposure notification apps are decentralized and privacy-preserving. Individual information is anonymous and stored on the smartphone of the user, rather than a centralized dataset, making the system more resilient to potential privacy breaches. However, data privacy is still a concern.

Privacy experts raised the concern that data could be re-identified by inference and when combining different data sources. Additionally, the Bluetooth and GPS function of the smartphone can be used by other apps to collect additional sets of information about the user that would not be subjected to the equal levels of protection and scrutiny that digital contact tracing and exposure notification apps have (Freeman).

**Limited information shared with public health authorities**

To preserve the privacy of the user, most digital tools have been designed to share little data with public health authorities. Exposure notification apps in particular give almost no information to public health authorities.

While this is a win for privacy, it also severely limits the ability of public health officials to use these tools to adjust their response to the pandemic. Additionally, this lack of data makes the effectiveness of digital contact tracing apps and exposure notification difficult to assess for researchers.

## Ethical Framework of Analysis

Taking from the European Convention on Human Rights, the International Covenant on Civil and Political Rights (ICCPR), and the United Nations Siracusa Principles (which sets the limitations of human rights principles at times of national and international crisis including public health emergencies), a team of Oxford researchers derived four guiding principles for the ethical deployment of digital contact tracing and exposure notification apps by national governments (Morley et al.). According to this framework, the technology has a green light for implementation if:

- there is evidence that it would improve public health outcomes (scientific validity),
- there are no better and less harmful alternatives (necessity),
- the deployment plan sets reasonable and realistic sunsetting provisions (time-boundedness)
- social or individual costs are justified by the gravity of the situation (proportionality).

Since these principles draw on broadly accepted ideas, they are going to base the analysis of the cases presented in the following sections: Ireland, Guayaquil (Ecuador), Haiti, and the Philippines. Those cases will outline how violating these standards would not only be socially harmful but would also hamper the effectiveness of the policy intervention by misaligning the incentives of the final users with the objectives of the policymaker.

**Scientific Validity: Evidence in the Field**

Epidemiological modeling and simulations by a team at Oxford University in April 2019, indicated that digital contact tracing could help slow down or stop coronavirus transmission (Oxford, 2019). Dr. Christophe Fraser, a co-author of the paper, who also served as Scientific Advisor to the UK Government Test & Trace Program and Group Leader in Pathogen Dynamics at Oxford University's Nuffield Department of Medicine, estimated that just 15% uptake of an exposure notifications system combined with manual contact tracing, could reduce infections by 15% and death by 11% in Washington State (MedRXiv, 2020).

On July 7[th,] 2020, Ireland became the first country to launch an exposure notification app, called COVID Tracker, which was developed by the software company NearForm. Within the first week, there had been 1.3 million downloads, which is approximately 37% of the population (Business Insider, 2020). On July 20[th], NearForm and Ireland's Health Service Executive (HSE) donated their app code base as open-source to the Linux Foundation, which would allow other jurisdictions to rapidly build their own apps (HSE July 2020).

By January 2021, several jurisdictions around the world had launched exposure notification apps, largely based on Google and Apple's Exposure Notification System (GAEN), with the NearForm-Linux Foundation partnership and PathCheck Foundation as the dominant providers. However, despite the adoption of the technology, much of the world was experiencing another wave of high infection rates which triggered another round of government-mandated lockdowns. This called into question the efficacy of exposure notification systems and the findings of the team from Oxford. A major problem was that the adoption rates in jurisdictions that had implemented the app were too low to be effective, though there may have been other contributing factors. In 12 of the U.S. States using the app, participation rates were in the single digits (New York Times, December 2021).

Another major challenge with the apps is the difficulty of integrating them into existing public health processes. Public Health practitioners require patient or population-level data which allows them to make rapid and informed decisions. During the manual contact tracing process, contact tracers will seek to extract location information from individuals who have tested positive for the coronavirus. They use this information to try to understand the spread of the virus in communities and where they may need to prioritize deploying resources such as test kits, which at the time, were in very limited supply. Google and Apple's exposure notification apps did not provide public health workers with that information, which was a point of contention for the two parties – Google and Apple vs. Public Health workers (Washington Post, 2020). The challenge for Apple and Google, of course, was finding the balance between public health utility and maintaining individual user privacy.

There are still unanswered questions on the efficacy of the apps. Defining efficacy is challenging and may be different in each context. Some public health authorities may define the efficacy of the apps as their ability to reduce the overall spread of infections, as compared to the manual contact tracing process. Others may define it within a more limited scope as the ability to identify and notify more people of possible exposures than manual contact tracers would. An empirical study looking at the effectiveness of Spain's RADAR COVID app was conducted by a team of researchers from the US, UK, and Spain. The team considered seven indicators that were integral to the epidemiological context: adoption, adherence, compliance, turnaround time, follow-up, overall detection rate, and hidden detection. The findings were that the detection rates using the app were 6.3 compared to just 3 for manual contact tracing (Nature, 2021)

The heterogeneity in the definition of efficacy for these digital tools combined with the diversion between what the results in the population are versus what initial studies concluded, makes it difficult to validate the utility of the tools. This, of course, presents a conflict between implementing to test and testing within a population without validating.

**Necessity: The Bottom-up Response in Guayaquil, Ecuador.**

By the end of March 2020, it was not uncommon to see coffins or corpses wrapped up in bedsheets on the streets of the city of Guayaquil, Ecuador (Mesa). The city, the economic capital of the country, had been severely hit by the COVID-19 pandemic, the first country in South America and the developing world to face the magnitude of the public health challenge.

In the first 15 days of April, an estimated 6,700 people died in Guayaquil, a city with an average pre-COVID mortality of roughly 1,000 people every 15 days (Mesa). At the time, the lack of widespread testing made it impossible to assess how many of them died of a SARS-CoV-2 infection. However, such a large deviation from pre-COVID numbers suggests that the high life toll was due to the pandemic, either the effect of direct infections or of the strained healthcare infrastructure that limited the response to other conditions.

In the middle of this healthcare crisis, a 32-year-old urban planner, Hector Hugo, was able to collect demographic and healthcare data provided by a friend at the Ministry of Health. He also accessed emergency call records that were briefly uploaded on a cloud system. From that data, Hector Hugo was able to filter calls requesting the collection of corpses and calls requesting support for severe respiratory symptoms (Dube and de Córdoba) at the time the only known symptoms of COVID-19.

Hugo used the data to create heatmaps of the infection, a visual tool that showed what areas of the city of Guayaquil were most severely hit by the pandemic. Thanks to the support of Carlos Bort, a Spanish data analyst who had worked on a similar project for the city of Madrid, the data collected were used to predict the areas of the city that were most vulnerable to the spread of the virus(Dube and de Córdoba)

While in the early months that effort was praised, it failed to receive political attention at the national level. The national response to the pandemic was slow and politicized, leading the Minister of Health to resign amid the health crisis (Mesa)

This data-driven approach was put to use in the city of Guayaquil first by Mesa Técnica, an interdisciplinary team that came together shortly after the first official COVID-19 case was confirmed at the end of February 2020 (Zúñiga) to support efforts to contain the spread of the Coronavirus(Moncada). Quickly after, the Mayor of Guayaquil, Cynthia Viteri, recognized the predictive tool developed by Hugo and Bort as an important asset to enable an effective public health response in the city.

Viteri put together the resources to bring the first line of the pandemic response out of the hospitals and into the streets of the Guayaquil. Health brigades(Dube and de Córdoba) and eight mobile clinics called "Unidades COVID" (Mesa) were sent off into the city but strategically concentrated in the most vulnerable neighborhoods. Poor overcrowded districts, present in Guayaquil as in most urban areas in the developing world, were areas of greatest concern.

By June 2020, the number of cases had massively decreased (Dube and de Córdoba) bringing the pandemic under control in the city while the crisis had shifted to other areas of the country (Human Rights Watch). Guayaquil was celebrated as the city that beat the Coronavirus. It is difficult to disentangle whether the drop in cases was an effect of herd immunity—a large fraction of the city had already contracted the virus—or of this innovative and proactive public health response. Most likely it was a combination of the two.

In a state of emergency, the effective allocation of scarce public health resources requires information. The approach developed by Hugo and Bort did not require personally identifiable information. Aggregate data on symptoms and infection rates by geographic areas, like neighborhoods, were enough for their analysis (Zúñiga). Only the inability to access those aggregated data led them to scout for individual health information from the leaked emergency calls to feed into heatmaps and predictive models of the spread of the virus.

The case of Guayaquil shows that alternatives to an approach that focuses on assessing the individual probability of infection by tracking people's movements, the approach of digital contact tracing, might be available, questioning its necessity. It also shows that when those less harmful systems are not explored effectively, people might adjust by independently adopting solutions that while well-meaning, might be more damaging to basic individual rights or liberties.

**Time-Boundedness: Transferring User Application Data in Haiti**

The country of Haiti has historically been plagued by economic and political volatility. Issues of corruption, disease outbreaks, and weak institutions have stagnated economic progress and the development of basic public infrastructure. This extends to the healthcare sector where a robust system for data gathering is non-existent, making it challenging for rapid and informed decision-making processes.

By April of 2020, in preparation for a potential widespread outbreak, the Haitian government and the country's health department, supported by the Organization of American States (OAS) – a 35-member multilateral organization formed to establish among its member states, an order of peace and justice, promote solidarity, strengthen collaboration, and to defend their sovereignty, their territorial integrity, and their independence – sought to implement a digital tool to aid with manual contact tracing. The decision to adopt this tool came from the realization that the country lacked enough trained contact tracers and given these low numbers, they could very quickly become overwhelmed should the prevalence of disease skyrocket.

Once that decision had been made, understanding the functionality of existing technologies and deciding on the one best suited for the Haitian context was the next step. Having weighed several variables including the level of data that would be most useful to contact tracers and privacy concerns, the parties concluded that a GPS-based tool was the best option and they entered into an agreement with PathCheck Foundation as the supplier.

The GPS tool developed by PathCheck Foundation is designed as a hybrid system which is different from the Bluetooth-reliant Google and Apple Exposure Notification (GAEN) in that they use different sensors and provide different levels of data. The GPS tool consisted of two components, a mobile app that residents would be asked to download called PathCheck GPS+ and an exposure notification tool, called "Safe Places" that would allow contact tracers to visualize areas of disease spread in communities and send out notifications to individuals who may have been in close contact with someone who tested positive for the virus. Details of the location, individuals with whom they may have been in close contact, or date and time information are not disclosed.

The rollout steps were multifold and included assessing the smartphone penetration rate in the country, potentially conflicting legislation that might prohibit implementation of such tools, cultural sensitivity and dynamics, economic conditions of households, and management of data. Several focus groups were conducted with community organizations and youth groups to understand cultural dynamics and sensitivity to the use of such tools, particularly because the tools were developed by an outside organization. There was also strong support from the private sector and a public-private partnership was entered into with Digicel Group, one of the two largest telecommunications providers in the Caribbean, to zero-rate the app—a practice of not counting use of an app or internet website against a user's data allotment. This was done to ensure that the use of the app would not be a financial burden on users and equitable distribution in access. For users who did not possess a phone with the capability to download an app, PathCheck Foundation sought to implement a paper-based, QR code solution called COVI-ID, developed at the University of Cape Town for the African contexts, in response to issues encountered in many developing countries, the prominent one being smartphone penetration.

By the summer of 2020, a limited number of the app had been downloaded in Haiti, especially among the group of trainees who had been brought on to conduct contact tracing, some of the youth and community groups with whom there had been engagements in the early stages of implementation, and perhaps other community members who had heard about the app. However, the nonexistent public health digital infrastructure into which the Safe Places platform could be integrated and a stunningly low incidence of the disease in Haiti, the project was deprioritized. Instead, the focus was shifted to transforming the country's health system. As of April 2021, out of the country's 11.49 million residents, there were 12,803 reported cases with 11,317 of them recovering and 252 deaths, according to data by Johns Hopkins.

With funding from the OAS, a new call center was built in Port-au-Prince to conduct contact tracing, disseminate information to residents via a hotline, and act as a conduit between the national testing lab and the Haitian Institute for Statistics and Informatics (IHSI), which had been appointed by the federal government to oversee all digital innovation projects. This newly established lateral system was designed to improve efficiency in the flow of information from the lab to residents, as well as to begin data collection by IHSI, which would then be used to inform public health decision-making.

Several issues arise with respect to the use of this data. First, Haiti has no clear legislative guidance on the interpretation or classification of data derived from apps as health data. In the case of the GPS app, the data provided is in the form of longitudinal and latitudinal points. Second, the use of the Safe Places tool is often accompanied by a sunset clause, automatically deleting user data after 14 days. However, if that data is then transmitted to IHSI's infrastructure, it would no longer be under the settings of the contact tracing tool and would violate the guiding principle of time-boundedness since the intent is to store that data over an extended period. There is also the matter of consent. A user of the app who chooses to share their location data for contact tracing purposes, would not be aware of how their data might be used beyond contact tracing efforts unless explicitly stated by the contact tracer.

While the intention to develop an optimally functioning public health system and maximize utility may be noble, transferring the use of individual data from one setting to another may violate their right to choose the purposes for which their data is used in the long run. As Haiti becomes more sophisticated with its government-controlled digital infrastructure and given its history of human rights violations, this scenario presents a potentially dangerous pathway to future abuse.

**Proportionality: Digital Contact Tracing in the Philippines.**

With a couple of hundred confirmed cases every day, in April 2020 (World Health Organization, "Philippines: WHO Coronavirus Disease (COVID-19) Dashboard With Vaccination Data"), the Philippines were still spared by the severity of the COVID-19 pandemic. But the Filipino government was conscious of the risk. At the time, Italy—a country with half of the population of the Philippines (OECD; World Bank)—had around 200 deaths every day and got to more than 4000 confirmed cases daily (World Health Organization, "Italy: WHO Coronavirus Disease (COVID-19) Dashboard With Vaccination Data"). The Philippines, like the rest of the world, was preparing for the challenge ahead.

Traditionally a role of the Department of Health, on April 12, 2020, the Filipino government entrusted contact tracing efforts to the Department of Defense (Gonzales). The months that followed saw the struggle to develop a national digital contact tracing app to control the pandemic. The company in charge of developing and launching StaySafe.Ph sought the support of PathCheck Foundation (PCF).

At the time, before the rise in popularity of Google and Apple's exposure notification system, blue tooth technology was considered too imprecise and burdensome on smartphone battery power to deliver an effective digital contact tracing solution. PCF's app, Safeplaces, was based on GPS technology. While still privacy-preserving with the appropriate accountability mechanisms, it allowed public health officials to generate heat maps of the spread of contagion to inform the use of scarce public health resources and notify individuals at high probability of contagion.

However, the adoption rate, stagnating at around 1%, was too low to be effective. Congress initiated an investigation into the StaySafe.Ph app over alleged privacy concerns in July 2020 (GMA News). The government commissioned the development of a new proposal to a military contractor who continued the relationship with PCF.

By that time, the number of confirmed COVID-19 cases was increasing quickly (World Health Organization, "Philippines: WHO Coronavirus Disease (COVID-19) Dashboard With Vaccination Data"), reaching quadruple digits. With the highest contagion rates of the ASEAN region, the Filipino government felt the need to expedite the launch of a digital contact tracing system. The plan was to continue the development of a contact tracing app based on GPS technology and concomitantly develop a blue tooth app based on Google and Apple's exposure notification system, which by this time was rising as a digital contact tracing standard.

Even though PCF's app was considered privacy-preserving, the use of GPS technology was controversial. For public health benefits to outweigh individual risks, it was necessary that the location data were voluntarily disclosed by the user to public health authorities and that the appropriate use of the data collected was ensured by well-designed accountability mechanisms. But new projected features of the Filipino digital contact tracing plan started to emerge.

The government contractor presented a new smartphone application: SaferPass, a digital ID. Based on the exposure notifications received by the blue tooth and GPS digital contact tracing apps, SaferPass was meant to generate an individual pass with a QR code, individual photo, and a timestamp. The pass was granting free movement to those with no exposure to COVID-19, and limit access to public spaces to those who had received an exposure notification via the smartphone applications or who had tested positive to COVID-19. It was a measure to enforce isolation and quarantine. Nevertheless, it was limiting individual freedom of movement even though the accuracy of exposure notification was still unknown for lack of data and questioned by many researchers.

For SaferPass to work, it required an ambitious monitoring infrastructure that included widely distributed automated kiosks, with scanners to read the QR codes generated by the SaferPass app. According to the plan presented, the kiosks were going to feature $360^O$ cameras with cloud video storage, microphones, AI-powered motion detection, and thermal scanners. For those individuals without a smartphone, the kiosks were meant to read physical QR code cards or communicate with 2G phones via text messages. Those allowed the government to keep a detailed track of people's movements in the absence of GPS data.

The potential privacy harm was disproportionate to the projected benefits, as the public health benefits of digital contact tracing solutions were and still are being debated. The massive monitoring infrastructure required for the system to work called into question its projected lifecycle. Unable to find an independent third party that could monitor the deployment of this technology and minimize risks, PCF quit this engagement.

## Conclusions

For many governments, implementing new technologies can be daunting and have far-reaching repercussions. If the technology does not perform as intended, it can slow down critical processes at best and cost lives, at worst. There could be unintended consequences with detrimental effects to local communities and can cost policymakers their jobs. There is no one method to implementing new technologies. In fact, a context-specific approach should almost always be adopted. Several considerations should be taken across a range of domains to minimize unintended harm. Governments should be fully transparent about the technologies they are implementing and encourage active discourse across the relevant sectors of society, assess the strength of institutional relations between residents and the government, with trust being a major component. They should consider all potential externalities from the implementation and ensure that the reasons for adopting the technology do not run afoul of human rights principles. Finally, governments should ascertain that the existing policies and regulations do not conflict with the activities for which the technology was developed.

Transparency is a fundamental element of any successful implementation of public goods. Community members should play active roles in the decision-making process and their concerns should be properly addressed. Governments must clearly communicate their intent to adopt new technologies, articulate the purposes for which they are being adopted, why these specific technologies are needed, how they will work to mitigate challenges, what the alternatives are, and who all the players are, both in the development and implementation processes. This includes the developers of the technology and any suppliers. Governments then need to create forums for residents to share their concerns and provide alternative ways that are anonymous, or those who wish to maintain their privacy. In a study by Harshini Jayaram on the adoption of critical digital public health applications in the USA, UK, and Israel (H Jayaram, 2021), survey respondents were asked to rank who they would trust most if asked to download a public health app. The study found that trust in government was lowest, among a group comprising religious leaders, non-profits, and academic institutions. This means that implementing technologies that require substantial adoption rates to be effective will require strategizing how to overcome the challenge of low trust.

Each sector within an economy may react differently to the implementation of new technologies. For example, policies mandating contact tracing apps be downloaded by residents of a country and extending that mandate to all arriving visitors could be good for the commercial sector, since it could minimize restrictions on the free movement of people within that country during a lockdown period, but bad for the tourism sector, if the majority of the arrivals are from countries where there is no mandate for using such apps. If tourism is among the largest contributors of revenue to that economy, workers in the travel and tourism industry may not be supportive of the implementation.

Perhaps most important when implementing new technologies is to ascertain that neither its function nor using it in ways outside the scope for which it was developed, create opportunities for violating human rights. Sunset clauses, restrictions on access, decentralization, and anonymization of data, privacy, and security are all elements that must be considered with appropriate steps taken to confirm compliance. This is of particular importance for technologies that have been developed in one country and distributed internationally. For example, using a GPS-based exposure notification application in St. Lucia is less worrisome than it would be if used in the Philippines. In St. Lucia, protecting user privacy was a primary concern of both the government and the department of health. They were intent on ensuring the security of the user data and restricting access to only approved individuals within the department of health. This was not the case in the Philippines where the government had empowered the military to oversee operations related to the implementation of contact tracing technologies and requested continuous access and monitoring of their citizens' location data, effectively setting up a surveillance state. The application was developed in the United States and was in alignment with commonly used security standards. Yet, the outcomes could be vastly different depending on the context.

There could be multiple negative externalities associated with the implementation of new technologies. Governments need to consider those may be and design strategies to mitigate their impact. In the case of exposure notification apps, it is imperative to think through possible unintended outcomes such as overconfidence. Community members who do not receive notifications via the app may believe that they are free from infection, but could be positive. This could increase community spread. Established public health services may be affected if residents are uncertain of the security of their health data or whether it will be used in nefarious ways. A single data breach could scare people from accessing important health services and change their views of health departments. Beyond public health, implementation and mandating that all residents use these digital solutions could increase the inequity gap. For example, in the case of "immunity passes" or "vaccine credentials," vulnerable groups like immigrants may be further disadvantaged if they fail to adopt these tools and are then denied access to critical services like education and finance. The consequences of this could be dire for entire communities.

Finally, existing policies should not conflict with the operations for which the technology is being implemented. For example, the health department in Panama wanted to implement an exposure notification app to minimize the burden on contact tracers, yet existing policies decreed that only licensed, practicing physicians could engage with residents who received notifications of possible exposures from the app. This policy conflicts with the intended purpose for which the technology was designed, in that it transferred the burden of executing procedures for a potential exposure from contact tracers to physicians, who were already overwhelmed by rising caseloads. Appropriate levels of training delivered to contact tracers to manage users who received exposure notifications is more efficient, especially given the number of false-positive notifications that are distributed.